\documentclass[conference,compsoc,letterpaper]{IEEEtran}
\pdfoutput=1

\ifCLASSOPTIONcompsoc
  \usepackage[nocompress]{cite}
\else
  \usepackage{cite}
\fi
\usepackage{algorithm}
\usepackage{algpseudocode}
\usepackage{tikz}
\usepackage{enumerate}
\usepackage{amssymb}
\usepackage{amsmath}
\usepackage{amsfonts}
\usepackage{graphicx}
\usepackage{multirow}% http://ctan.org/pkg/multirow
\usepackage{hhline}
\ifCLASSINFOpdf
\else
\fi
\begin{document}
\title{Towards a Complete Framework for Virtual Data Center Embedding}

\author{\IEEEauthorblockN{M. P. Gilesh}
\IEEEauthorblockA{National Institute of Technology Calicut\\
Kozhikode, Kerala, India\\
Email: gilesh\_p140059cs@nitc.ac.in}}
\maketitle

\begin{abstract}
Cloud computing is widely adopted by corporate as well as retail customers to reduce the upfront cost of establishing computing infrastructure. 
However, switching to the cloud based services poses a multitude of questions, both for customers and for data center owners. 
In this work, we propose an algorithm for optimal placement of multiple virtual data centers on a physical 
data center. Our algorithm has two modes of operation - an online mode and a batch mode. 
Coordinated batch and online embedding algorithms are used to maximize resource usage while fulfilling the QoS demands.
Experimental evaluation of our algorithms show that acceptance rate is high - implying higher profit to infrastructure provider. 
Additionaly, we try to keep a check on the number of VM migrations, which can increase operational cost and thus lead to service 
level agreement violations. 
\end{abstract}


% \fi
\IEEEpeerreviewmaketitle

% \keywords{Cloud Computing; Data Center Virtualization; Network Virtualization; Virtual Data Center Embedding.}

\section{Introduction}
In the last few years, cloud computing has become the backend of all the service oriented IT businesses. 
The conventional service providers (SP) are able to significantly reduce their capital expenditure (CapEx) by switching to virtualized IT infrastructure.  
The time to establish (ToE) infrastructures of any desired size has come down from several months to a few minutes/hours of automated provisioning of virtual machines (VMs) in 
physical infrastrcture providers' premises and installing the application stack on these virtual infrastructures \cite{demchenko:2016} using tools like Puppet, Ansible, and Vagrant. 

With the new model of service provisioning, the business role of conventional SPs has got divided among infrastructure providers(InP), 
who expose a software abstraction of the underlying hardware or software resources; and service providers, who use these abstractions to build their own application stack to offer services to the end users.
Virtualization of the hardware is the prime technology that enabled the cloud computing paradigm. Though the concept of virtualization is more than two decades old, 
cloud computing has leveraged the expoitation of the potential of virtualization at various levels viz. hardware, development platform and software. 
As of now, almost every domain of computing infrastructure, including computing, storage, network, operating systems and applications, have been virtualized to maximize the resource utilization and 
reduce the ToE. The InPs maintain a pool of resources to be abstracted and mutliple instances of user machines/ applications run on these resources, typically called \textit{slices}.

Network virtualization, though late entrant in cloud computing,  has key role in complying with the SLA between InP and SPs.
The bandwidth requirement for internal traffic is growing fast against that for external traffic in cloud data centers\cite{greenberg:2009}.
From the business perspective, virtualizing the network is crucial because  the cost of networking is 
escalating against the cost of other equipments and the ratio of network to compute is going up, especially in big data applications.
Literature show that, NV is killer application for the software defined networks (SDN) technology. 
In this paper, we address the problems associated with sharing compute and network resources of an InP, in particular,  that of a datacenter infrastructure.
In multi-tenant data centers, multiple service deploy their virtual infrastructure(VI) for service provisioning. The service level agreement (SLA) defines the QoS expectations of the service providers
and the penalty for the violation of the same. Hence, it is critical for the InP to provide isolation between these VIs, in terms of performance, disruption, and security.

In this paper we address the virtual data center embedding (VDCE) problem in cloud data center environment\cite{correa:2015}. 
In VDCE, embedding of virtual networks (VNs) and virtual machines (VM) are considered simultaneously. 
A virtual datacenter(VDC), like any other DC, has VMs connected through virtual switches using virtual links of specific bandwidth and delay constraints.
VDCE refers to the class of algorithms which optimally superimpose multiple VDCs on a DC, satisfying the SLA. \cite{wang:2016,rosa:2014}. 
The problem is important because, many customers (i.e. SPs) try to create or migrate their topology intact (including switches) to a cloud for reasons like failure recovery, scaling or economic saving.
% The PortLand\cite{mysore:2009}, VL2\cite{greenberg:2009}, SecondNet\cite{guo:2010}, CloudNaaS\cite{benson:2011} etc. are some of the solutions 
% in literature to address the practical difficulties in superimposing VNs on an SN.  
% Bari et al. \cite{bari:2013} give a detailed survey on the well known implementations.
% Optimal placement of multiple VNs on their physical counterpart to satisfy the objectives of InP and/or SP is a well studied problem.

\begin{figure}[h]
  \centering
   \includegraphics[width=.45\textwidth]{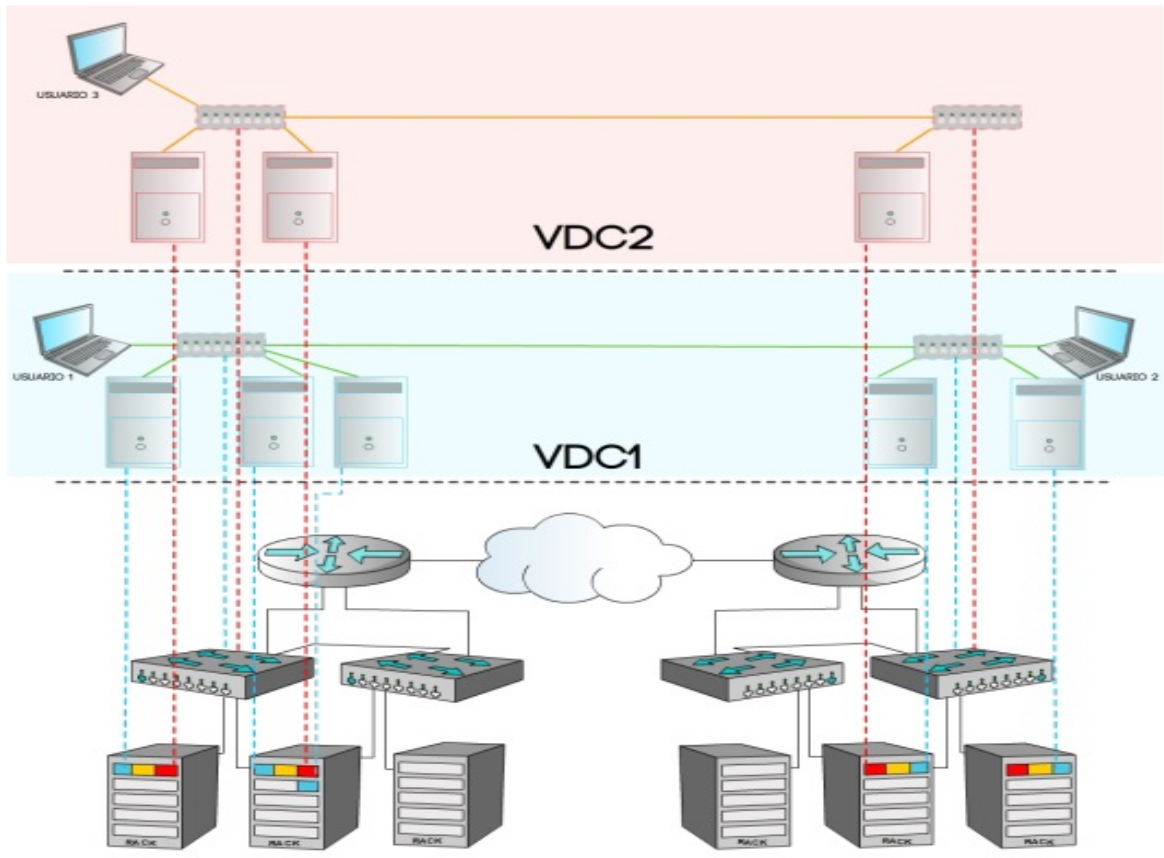}
  \caption{Virtual Data Center Embedding \cite{correa:2015}}
  \label{fig:vne}
\end{figure}

Formaly, the problem of  VDCE, is defined as follows. Let $G^s = (N^s, E^s, C^s, \phi^s_{a}, \delta^s_b, \xi^s_c)$ be a physical DC and there is a set of VDC requests
$ \mathcal{R}=  \{ G^v_u = (N^v_u, E^v_u, C^v_u, \phi^v_{a_u}, \delta^v_{b_u},  \xi^v_{c_u})$ : $u=1,2,..m\}$  .   
% Table 1 gives the description of the variables used. 
Table~\ref{tab:var} gives the description of the variables used. 
The problem is to find an optimal placement of VDCs on DC while satisfying the SLAs with the tenants $u=1,2,..m$. 
In otherwords, find a mapping from VDC set to DC 
% $$ \mathcal{F_i} : G^v_i \mapsto ({N_s}', {\mathcal{P}_s}', {\delta^s}', {C^s_p}') $$
 $$ \mathcal{F}^C_u : {C^v_u} \rightarrow {C^s}'$$
 $$ \mathcal{F}^N_u : {N^v_u} \rightarrow {N^s}' $$
 $$ \mathcal{F}^E_u : {E^v_u} \rightarrow {\mathcal{P}^s}' $$ for $u=1,2,..m$ where ${\mathcal{P}^s}' \subset {(E^{s})}^k, 1\le k\le |N^s|, \,\,\, {N^s}' \subseteq N^s , {C^s}' \subseteq C^s$, 
under the following conditions 
\begin{enumerate}
 \item Virtual machines(\textit{VM}) are mapped to servers/physical machines (\textit{PM}).
 \begin{equation}
   \mathcal{F}^C_u(c_1) = c \in {C^s}'
  \end{equation}
 
 \item Not more than one node of a virtual request is mapped to a physical (substrate) switch (\textit{pSwitch}).  
 \begin{equation}
   \mathcal{F}^N_u(n_1) = \mathcal{F}^N_u(n_2) \implies n_1 = n_2
 \end{equation}
  and  hence, at most one virtual link (\textit{vLink})  of a request is mapped to a path. 
  Moreover, virtual edge switches  are mapped only to edge switches of the physical topology.
 
 \item Sum of capacity demands of \textit{VM}s  does not exceed the total capacity of the \textit{PM} to which they are mapped
 \begin{align}
  \sum_{u \in \Re: \mathcal{F}^C_u (k) = s}  \phi^v_{a_u}(k) \le \phi^s_a(s)\,\,\,\,\,\,\,\,\,  %\nonumber \\ 
   \forall s \in C^s, \forall a  \in q^C
 \end{align}
 
 \item Similarly, sum of capacity demands of virtual switches (\textit{vSwitch}) should be within that of the \textit{pSwitch} to which they are mapped
 \begin{align}
  \sum_{u \in \Re: \mathcal{F}^N_u (k) = n}  \delta^v_{b_u}(k) \le \delta^s_b(n)\,\,\,\,\,\,\,\,\,  %\nonumber \\ 
   \forall n \in N^s, \forall b  \in q^N
 \end{align}
  
 \item The sum of attribute demands of \textit{vLinks} do not exceed the total capacity of the physical / substrate link (\textit{pLink} ) to which they are mapped.
 \begin{align} 
  \sum_{u \in \Re:\mathcal{F}^E_u (l)=p; e\in p; p\in \mathcal{P}^s}  \xi^v_{c_u}(l) \le \xi^s_c(e) \,\, \nonumber \\
  \forall e \in E^s, \forall c \in q^E
 \end{align}
 \end{enumerate} 

Figure~\ref{fig:vne} illustrates how two VDC requests are embedded on a physical data center.

\begin{table}
\begin{center}
 \begin{tabular}{p{.04\textwidth}|p{.39\textwidth}} \hline \hline
  \textbf{Var} & \textbf{Description} \\ \hline
  $\Re$  & Set of serviceable VDC requests\\
  $N^s$  &Set of substrate switches \\
  $E^s$  &Set of substrate links \\
  $C^s$  &Set of servers \\
  $\mathcal{P}^s$  & Set of all paths in substrate network\\
  $N^v_u$  &Set of virtual switch of request $u$\\
  $E^v_u$  &Set of virtual links of request $u$\\
  $C^v_u$  &Set of VMs of the request $u$\\
  $\phi^s_{a} / \delta^v_{a_u}$ & Value of \textit{PM /VM} attribute $a$ \\
  $\delta^s_{b} / \delta^v_{b_u}$ & Value of \textit{PSwitch / VSwitch} attribute $b$ \\
  $\xi^s_{c} / \xi^v_{c_u}$ & Value of \textit{PLink / VLink} attribute $c$\\
  $q^C $  & Set of machine attributes \\
  $q^N /q^E $  & Set of switch/machine/link attributes \\
  $\mathcal{P}_{kl}$  & Set  of paths in SN between $k$ and $l$ \\
%   $X_{c}$ & - & 1 if a configuration $c$ of a VNR is mapped, Else 0 \\
  $y^u_{ij,kln}$  & 1 if vLink $ij$ of  $u\in \Re$ is mapped to \textit{n}th path $kl$.\\
  $x^u_{ik}$  & 1 if vSwitch $i$ of $u\in \Re$ is mapped to a pSwitch $k$, Else 0 \\
  $w^u_{ik}$  & 1 if VM $i$ of $u\in \Re$ is mapped to a PM $k$, Else 0 \\
  $Z_u$  & 1 if a request $u$ is mapped. Else 0.\\
  $p^{e}_{kln}$  & 1 if $e$ is in the \textit{n}th path of $kl$. Else 0 \\
  \hline
 \end{tabular}
 \end{center}
 \caption{Notations used}
 \label{tab:var}
\end{table}

The node mapping problem - mapping virtual nodes to physical nodes - was proven to be NP-Hard \cite{anderson:2002}\cite{chowdhury:2010,Yu:2008}. 
The problem is reducible from the famous Multiway Separator Problem\cite{anderson:2002} which is NP-Hard. 
Optimal assignment of links with functional constraints, in a graph, where the nodes are already assigned, is still NP-hard \cite{chowdhury:2010,Yu:2008}
and is similar to Unsplittable Flow Problem (UFP)\cite{haider:2009} / Multi-Commodity Flow (MCF) problems. Many heuristic solutions were proposed for the problem.

We propose a VDCE algorithm that employs suitable technique for embedding requests depending on the  availability of DC resources.
The model employs mathematical programming for embedding a group of requests, in one shot. 
Local search heuristic is used for online embedding of requests to reduce the average waiting time of the requests. 
The online embedding reduces the number of VM migrations by not consolidating the VMs, as proposed by many solutions found in literature. 
% A local search solution is proposed to keep the resulting virtual machine migrations under control.

Rest of this paper is organized as follows. Section~\ref{sec:related} gives a brief review of the relevant work in this area. Section~\ref{sec:proposed} explains the proposed algorithm in detail. 
In Section~\ref{sec:results}  we discuss experimental setup for evaluation of our algorithm and the results of these experiments.
Section~\ref{sec:conclusion} concludes the paper.

\section{Related Work} \label{sec:related}
Virtual Network Embedding problem, for efficient placement of virtual networks on physical network, is being studied for several years .
Algorithm to efficiently map virtual nodes to the physical nodes, to satisfy various objectives had been studied. 
Single shot embedding of both nodes and edges were also studied \cite{Belbekkouche2012}\cite{fischer:2013}. 
In practice, exact algorithms for the problem apply for small networks only. 
A comprehensive survey of classic virtual network embedding techniques is given by Belbekkouche et al. \cite{Belbekkouche2012} and Fischer et al. \cite{fischer:2013}.
Fischer et al. \cite{fischer:2013} proposed a taxonomy for VNE and classified them as centralized vs distributed, static vs dynamic and redundant vs concise.
The centralized, dynamic and redundant algorithms are particularly of interest because they apply to the present day data center networks, especially the software defined networks

The VDCE problem is different from VNE because the latter cares about 
the efficient palcement of VMs with multiple constraints. Efficient placement of VMs on servers is an independent problem by itself.
To the best of our knowledge, there are only a few proposals for VDC resource allocation in cloud 
computing centers. Sivaranjini et al. \cite{sivaranjini:2015} and Correa et al. \cite{correa:2015} give surveys on 
the existing virtual data center embedding algorithms. 
Papagianni et al. \cite{papagianni:2013} proposed a VNE for cloud computing environment viz. Networked Cloud Mapping (NCM), that handle both 
compute and network node assignments. The mixed integer programming (MIP) model attaches multiple property vectors for the nodes including the type 
of nodes, capacity etc.

Zhani et al. \cite{zhani:2013} proposed a VDC planner to embed virtual data center network and machines on a substrate data center. The primary consideration of the proposal is migration-awareness. There
are three scenarios considered in the proposal -  an initial deployment, handling up/down scaling of VDC and a dynamic consolidation (reoptimzation) for power saving. The experimentations are carried out on  
the VL2 topology\cite{greenberg:2009}. 
Three stage virtual data center embedding algorithm by Rabbani et al. \cite{rabbani:2013} maps virtual machines, switches and links in sequential stages. The heuristic algorithm reduces the server fragmentation, 
communication cost and the resource utilization. 

Amokrane et al. \cite{amokrane:2013} give a resource allocation technique for virtual data center spanning over distributed substrate data centers. 
The proposed method has two stages - a VDC partitioning and a partition embedding. In partitioning, the VDC requests are partitioned to minimize 
the inter-data center bandwidth. In the second phase, the partitions are mapped to data centers satisfying capacity constraints. 
The integer linear program (ILP) formulation of both phases ensures exact solution for reduced bandwidth.
Venice is a reliability aware VDC embedding algorithm proposed by Zhang et al. \cite{qizhang:2014}. The authors proved that computing the availability of VDC is a hard problem. The reliability-aware
embedding is handled using a heuristics and a consolidation to handle the frequent entry and exit of requests. 
Wang et al. \cite{wang:2016} proposed a heuristic framework for the  VDCNE problem - \textit{Presto}. The framework uses Blocking Island (BI) paradigm for improving the accuracy and speed of embedding. However,
the link and node embeddings are uncoordinated and the heuristic based solutions are sub-optimal.
% {\centering
% \begin{table}
%  { 
%  \tiny
%  \begin{tabular}{|l|p{1cm}|c|p{1.2cm}|p{1.2cm}|} \hline
%   \textbf{Proposal} &  \textbf{Type of Solution} &  \textbf{Optimality} &  \textbf{Acceptance Rate} &  \textbf{Embedding Speed} \\ \hline \hline
%   Papagianni\cite{papagianni:2013} & MIP & Optimal &High & Slow \\ \hline
%   Amokrane\cite{amokrane:2013} & ILP & Optimal & High & Slow \\ \hline
%   Zhang\cite{qizhang:2014} & Heuristic & Sub-optimal &Medium & Fast \\ \hline
%   Rabbabi\cite{rabbani:2013} & Heuristic & Sub-optimal & Medium & Fast \\ \hline
%   Zhani\cite{zhani:2013} & Heuristic & Sub-optimal & Medium & Fast \\ \hline
%   Soares\cite{soares:2014} & ILP & Optimal & High & Slow \\ \hline
%   Wang\cite{wang:2016} & Heuristic & Sub-optimal & Medium & Fast \\ \hline
%   \end{tabular} }
% \caption{Comparison of VCDE/VDCNE solutions}
% \label{tab:compare} 
% \end{table}
% }

% Table~\ref{tab:compare} compares the existing VDCE algorithms in terms of the acceptance rate and speed of embedding. 
% The comparison 
Our survey shows that algorithms are either exact algorithms that suffer from sluggishness or heuristic algorithms which leave the DCs under-provisioned. 
A suboptimal embedding might be possible using heuristics. But, such an embedding may lead to rejection of other peer requests, resulting poor QoE (Quality of Experience) of customers. 
Moreover, most of the heuristics fail if the residual resources are spread over multiple fragments. 
Hazzles of live VM migration dismiss the possibility of frequent remapping of the incumbent nodes. There had been solutions which use VM migrations to find optimality, neglecting the cost of migration.  
So, we believe that, a well coordinated hybrid technique can find an optimal solution in less time. We propose a mixed approach with 
suitable modes from mathematical programming and local search heuristics to maximize the resource usage with limited VM migrations.

\section{Proposed Algorithm} \label{sec:proposed}

% The proposed HyViDE framework has three stages working in coordination. 
% \begin{enumerate}
%   \item \textit{Admission Control} : Provisionally selects a set of serviceable requests to favor the objective(s) and populate the set $\Re$. 
%   The aim is to reduce the number of requests to be handled in static embedding and also to prioritize the requests in dynamic embedding.
%  \item \textit{Static/Batch Embedding} : Embed a batch of selected requests from $\Re$ with any batch embedding algorithm. 
%  We use an ILP model for finding an exact solution.
%  \item \textit{Dynamic Embedding} : Embed a pending request or a new request on arrival, and a local modification of the embedding is 
%  required to recover from a substrate failure or a virtual network requires a scaling.
% \end{enumerate}

% The SPs in multi-tenant virtualized data centers have the flexibility to manage their network in isolation, without the intervention of the InP. 
% The tenant can establish their own mechanisms to improve fault tolerance 
% (survivability) of the virtual network. Hence, we believe that path splitting\cite{Yu:2008} is not suitable for a fully virtualized network environment. 
% Moreover, path splitting will affect the performance of real time video applications and several other services, including those that rely on jumbo packets.
% Due to the performance considertations of coordinated algorithms in three stages, our strategies in HyViDE do not use path splitting or resource reservation for failure recovery.
The proposed algorithm works in two different modes - online mode or batch mode - based on the rate of fragmentation of resources with in data center.
Different techniques are used to perform batch embedding and online embedding.
If suffiecient resources are available, individual arriving requests are embedded immediately with local search method to avoid reoptimization of the entire mapping. 
Online embedding is done also when one or more embedded requests exit and if the total residual resources are above a threshold ($t^1$) defined in terms of the smallest pending request   .
Online embdding is attempted in case of failure of hardware components also. 
Frequent reoptimization of the entire embedding or consolidation may cause many VM migrations. 
It may cause unprecedented delay in the communication within a virtual network, 
if a tenant \textit{vSwitch}  is mapped to a farther \textit{pSwitch} after reoptimization. 
Batch embedding of multiple requests is performed by a mixed integer programming model. Batch embedding is performed if there are sufficient resources to embed multiple VDC requests or the 
total fragmented resources are above a threshold vector $t^2$ defined in terms of the resource requirement of the largest pending request. 
Threshold vector has values $t^2_c, t^2_n, t^2_e$ the residual values of servers, switches and links.

Initially, the residual resource vectors $t_c$, $ts_n$, $ts_e$ are initialized with the capacity vectors of the data center. 
If there are multiple pending requests and the total demand is less than $t_c$, $ts_n$, $ts_e$ we try a batch embeding.  
If the sum of resource demand vectors of all requests exceed  $ts_n$, $ts_e$ for nodes and links respectively, and multiple requests can be embedded, then a batch embedding is attempted on a selected 
subset request. The subset has the high priority jobs. Deatils of the online embedding and batch embedding algorithms are give below.

\subsection{Online Embedding}
Online embedding is the most common occurence in our model. In online mode, an embedding is attemped whenever arrival of a new request or an exit of embedded request happens.
Specifically, an online embedding is attempted when:
\begin{enumerate}[(i)]
 \item a new request arrives and the local remapping can enable the embedding of the request; or.  
 \item there is failure of some servers, nodes and links and VDC components already mapped to the failed part of the substrate requires a remapping; or  
 \item a mapped virtual data center requires scaling up and free resources are not available in the neighbourhood.  
\end{enumerate}
online embedding attempts to find a local solution with minimum modification to the existing mappings.
If there are multiple waiting requests, we prioritize the  requests based on the time to expiry, size of the VDC and expected time of incumbancy.
If a fragment (connected component) of the DC graph has sufficient resources to embed the selected request then, an embedding is attempted by simple heuristics.
We use a local search and swap technique to do online embedding. 
A temporary mapping of the new request is made  in the following order - VMs, Switches, Links. 
The temporary mapping is allowed  to have capacity violations. 
Then swapping is attemped between already mapped VMs/Switches in locations where the temporary mapping has least violations.
If a solution is derived by few number of swappings (limited by the number of swaps required) the temporary mapping is made permanent. 
If neither of the above are possible the algorithm tries to embed the next request.

% Algorithm~\ref{alg:vne} describes the working of HyViDE. 
% The resource vectors $ts_n$, $ts_e$ are set to the sum of residual resources of each type for all nodes. 

% In case of online embedding,  if the residual resources are higher than the upper threshold $(t^2_n,t^2_e)$, a batch embedding is attempted. If the residual resources are 
% between the lower and upper thresholds, a online embedding is attempted. The time to expiry of the request should be higher than 
% the sum of the current time and time required for deployment (TD). Otherwise no embedding is attempted.

\subsection{Batch Embedding} \label{sec:batch}

Batch embedding is attempted when a set of requests are available for the first time or when there are enough fragmented resources for embedding a new/deffered request for which 
it is hard to find a local solution. In our algorithm we focus on maximizing the acceptance rate of requests which positivley affects the revenue of InP and availability of the cloud service.
The batch embedding algorithm works as follows. 

% Similar to other flow based models\cite{hu:2013}, we create an auxillary graph as shown in Figure~\ref{fig:aux}.
% 
\begin{figure}[h]
  \centering
  \includegraphics[width=.45\textwidth]{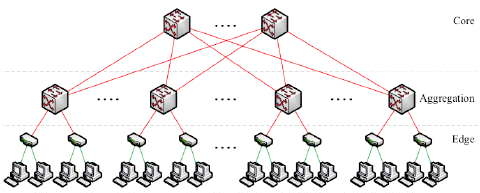}
  \caption{Data  Center Topology}
  \label{fig:fattree}
\end{figure}

% \begin{figure}[h]
%   \centering
% % %   \includegraphics[width=.45\textwidth]{fattree.png}
% %    \includegraphics[width=.45\textwidth]{fat_tree1.jpg}
% \begin{tikzpicture}[thick, node distance=1em,scale=0.35]
% 	% Level 1 switches
% 	\foreach \n in {0,...,1}{
% 		\node (level 1 \n) at (4.0cm + \n*4*3cm,9.0) [draw,rectangle,inner sep=0.6em] {{\tiny Switch}};
% 	}
% 	% Level 2 switches
% 	\foreach \n in {0,...,3}{
% 		\node (level 2 \n) at (2cm + \n*2*3cm,6) [draw,rectangle,inner sep=0.4em] {{\tiny Switch}};
% 		
% 		\draw [thick] (level 2 \n) to (level 1 0);
% 		\draw [thick] (level 2 \n) to (level 1 1);
% 	}
% 	\foreach \n in {0,...,7}{
% 		\node (level 3 \n) at (0.75cm + \n*2*1.5cm,2) [draw,rectangle,inner sep=0.2em] {{\tiny Switch}};
% 		
% 		\draw [thick] (level 3 \n) to (level 2 0);
% 		\draw [thick] (level 3 \n) to (level 2 1);
% 		\draw [thick] (level 3 \n) to (level 2 2);
% 		\draw [thick] (level 3 \n) to (level 2 3);
% 	}
% 	% Nodes
% 	\foreach \n in {0,...,15}{
% 		\node (node \n) at (\n*1.5cm,0) [draw,circle,inner sep=0.1em] {{\tiny PM}};
% 		
% 		\pgfmathtruncatemacro{\switch}{\n/2}
% 		\draw [thin] (node \n) to (level 3 \switch);
% 	}
% \end{tikzpicture}
%   \caption{Common Data Center Topology}
%   \label{fig:fattree}
% \end{figure}

We formulated the batch embedding problem as a mixed integer program (MIP). 
The model restricts at most one \textit{VSwitch} of a request on an \textit{PSwitch}.
However, this restriction does not apply to embedding of \textit{VM}s on \textit{PM}s. Multiple VMs can be embedded in a single PM. 
An edge switch of a VDC request in embedded on an edge switch of the substrate topology (Figure~\ref{fig:fattree}). 
Any \textit{VLink} is embedded on a single path between the nodes onto which the end virtual devices are mapped.
The MIP model for batch embedding is as follows
% VDC requests may have specific demands on other QoS parameters like latency, location, distance between edge nodes etc. 
% Our model applies to those requests which are in need of such special constraints. 

Equation~(\ref{eq:obj}) tries to maximize the number VDCs embedded, while minimizing the number of VM migrations, 
subject to the constraints (\ref{eq:node1}) - (\ref{eq:bound})
\begin{equation}
Maximize \sum_{u\in \Re}  Z_u  - r_1 - \frac{r_2}{f} \label{eq:obj}
\end{equation}
where $r_1$ and $r_2$ are the normalized migration distances of \textit{VM}s and \textit{vSwitch}es already embedded.
Parameter $f$ decides the weight of penalty for vSwitch migration relative to VM migration.
 
Constraint (\ref{eq:node1}) and (\ref{eq:vm1})ensures that when a VDC $u$ is embedded all its \textit{vSwitches} and \textit{VMs} are also embedded.
\begin{align}
 \sum_{k\in N^s} x^u_{ik}  = Z_u  \,\,\,\,\,\,\,\,\,\,\, \forall i \in N^v_u, u\in \Re  \label{eq:node1} \\
 \sum_{k\in C^s} w^u_{ik}  = Z_u  \,\,\,\,\,\,\,\,\,\,\, \forall i \in C^v_u, u\in \Re  \label{eq:vm1} 
\end{align}
As per the definition, at most one \textit{vSwitch}  of virtual request in embedded on a \textit{pSwitch}.  Constraint~(\ref{eq:node2}) affirms this rule.
\begin{equation}
\sum_{i\in N^v_u} x^u_{ik}  \le 1  \,\,\,\,\,\,\,\,\,\,\, \forall k \in N^s, u\in \Re  \label{eq:node2}
\end{equation}
To ensure that a virtual link is embedded only once and not split across multiple path, constraint~(\ref{eq:links}) is used. There can be exponentialy many paths between any two nodes 
of a graph. 
% However, in a fat tree topology shown in Figure~\ref{fig:fattree}, there are at most 4 simple paths between the nodes in different pods. Though we can find paths 
We consider paths of length utmost 4 and other paths are not included in $\mathcal{P}_{kl}$. 
\begin{equation}
\sum_{kl \in \{N^s \cup C^s\} \times \{N^s \cup C^s\}} \sum_{n\in \mathcal{P}_{kl}} y^u_{ij,kln}   =Z_u  \,\,\,\,\,\,\,\,\,\,\, \forall ij\in E^v_u, u\in \Re \label{eq:links}
\end{equation}
The linking constraint~(\ref{eq:nodedge}) ensures that, whenever a pair of \textit{vSwitches} are mapped, the \textit{vLink} between them are mapped to at most one path between the selected \textit{pSwitches}. 

\begin{equation}
\sum_{n\in \mathcal{P}_{kl}} y^u_{ij,kln} - x^u_{ik}\,\,\, x^u_{jl}  = 0  \,\,\,\,\,\,\,\,\,\,\, \forall ij\in E^v_u, u\in \Re \label{eq:nodedge} \\
\end{equation}
For any server in the DC, there is one (Figure~\ref{fig:fattree}) or a constant number of switches in the neighbourhood. The topology in use has only one neighbour. Hence the constraint
for embedding a (\textit{vSwitch}, \textit{VM}) link is given as 
\begin{align}
 \sum_{kl \in E^s\cap(N^s \times C^s)} y^u_{ij,kl1} - x^u_{jl} \,\,\, w^u_{ik}  = 0  \nonumber \\ 
 \forall ij\in E^v_u \cap (N^u_v \times C^u_v), u\in \Re \label{eq:edgelink}
\end{align}
The constraint (\ref{eq:servcap}) assures that the sum of resource demands of \textit{VMs} mapped to a \textit{PM} does not exceed its corresponding resource capacity.  
\begin{equation}
 \sum_{u \in \Re} \sum_{i \in C^v_u}  w^u_{ik}\,\,\, \phi_a (i) \le  \phi_a (k) \,\,\,\,\,\,\,\, \forall k \in C^s, \forall a \in |q^C|  \label{eq:servcap} 
\end{equation}

Similarly, constraint (\ref{eq:nodecap}) ensures that the sum of resource demands of \textit{vSwitch}es mapped to a \textit{pSwitch} does not exceed its corresponding resource capacity.  
\begin{equation}
 \sum_{u \in \Re} \sum_{i \in N^v_u}  x^u_{ik}\,\,\, \delta_b (i) \le  \delta_b (k) \,\,\,\,\,\,\,\, \forall k \in N^s, \forall b \in |q^N|  \label{eq:nodecap} 
\end{equation}
Similar to the above the sum of link resource demands of all \textit{vLink}s,  mapped to a \textit{pLink}, should be less than the capacity of the substrate.  
\begin{align}
  \sum_{u \in \Re} \sum_{ij \in E^v_u}  y^u_{ij,kln} \,\,\, p^e_{kln} \,\,\, \xi_c (ij) \le  \xi_c (e), \nonumber \\  
  \forall e \in E^s, \forall c \in |q^E|  \label{eq:linkcap} 
\end{align}
Following constraint put bounds on values that the variables can assume.
\begin{equation}
 Z_u \in \{0,1\},  y^u_{ij,kln} \in \{0,1\},  x^u_{ik} \in \{0,1\}, w^u_{ik} \in \{0,1\} \label{eq:bound}
\end{equation}

Apart form the constraints given above, VDC requests may have specific demands on other QoS parameters like latency, location, distance between VMs etc. The following set 
of constraints apply to those requests which are in need of such special consideration. If the VDC requests $u \in \Re ' \subset \Re$ want to limit the maximum latency between adjacent nodes
(switch/VM) to $d_u$,
the following constraint~(\ref{eq:latency}) is added in respect of them.
\begin{equation}
  \left( \sum_{vw\in P_{kln}} d_{vw} \right) y^u_{ij,kln} <= d_u \,\,\,\, \forall ij\in E^v_u ,\,\, \forall u \in \Re ' \subset \Re \label{eq:latency}
\end{equation}
where $d_{vw}$ is the delay of a physical link $(v,w)$

A vital requirement of VDC placement in cloud environment is the freedom to specify the location for embedding a VM. Customarily, the need
arises from the locality of other nodes, ease of access etc. This is a major requirement in virtual clusters performing big data processing, to
improve computation time. The following constraint is used to specify the exact substrate node $k$ on which a virtual node $i$ should be embedded. 
\begin{equation}
w^u_{ik} = Z_u
\end{equation}
A more flexible method would be specify a possible subset of \textit{PM}s to embed a given \textit{VM}. 
Optional constraint (\ref{eq:localityn}) limits the embedding VM $i$ on a \textit{PM} from a set of servers $C^{s'} \subset C^s$
\begin{equation}
 \sum_{k\in C^{s'}} w^u_{ik} = Z_u \label{eq:localityn} 
\end{equation}
% Those constraints in (\ref{eq:node1}) which become redundant, by the addition of these optional constraints, were removed to speed up the convergence.

As and when needed, batch embedding is done for compaction of the unusable fragmented resources. Such a re-embedding of incoming and 
mapped VDC request is VM migration aware and 
hence tries to minimize/limit the cost of changes in the existing mapping.  
We assume that virtual machines of different sizes have varying cost of migration. Hence costly migrations are avoided to the extent possible.   
If remapping a \textit{VSwitch} is necessary, the new \textit{PSwitch}, to which the mapping is done, 
is not far from the already mapped \textit{PSwitch}. This feature keeps a check on the number of VM migration and the extra internal traffic.

% Calculation of threshold values are as described in Section~\ref{sec:proposed}.
\begin{table}[h]
 \begin{tabular}{|p{.32\textwidth}|p{.1\textwidth}|}  \hline
  \textbf{Variable}  & \textbf{Values} \\ \hline 
  PM CPU Cores & 8 \\ \hline
  PM Memory & 16384 \\ \hline
  pSwitch memory  &  100 \\ \hline
  pLink bandwidth (core - aggregation)  & 10000 \\ \hline
  pLink bandwidth (edge - aggregation)  & 1000 \\ \hline
  pLink bandwidth (edge - server)  & 1000 \\ \hline
  No of VMs  & 40 - 100   \\ \hline
  VM cores & 1 - 2 \\ \hline
  VM Memory & 256 -512 \\ \hline
  No of vSwitches  & 5 - 20 \\ \hline
  vSwitch memory  & 10 - 25 \\ \hline
  vLink  bandwidth  & 5 - 200 \\ \hline 
  Duration of incumbancy  & 10 - 90 \\ \hline
 \end{tabular} 
 \caption{Simulation parameters}
 \label{tab:simvar}
\end{table}

\section{Preliminary Results} \label{sec:results}
% \section{SIMULATION} \label{sec:experiment}
We developed the simulation environment in python, with FNSS toolchain \cite{fnss} and NetworkX library in the backend, 
for managing the substrate and virtual networks . 
The batch embedding is implemented with a commercial solver, CPLEX\cite{cplex}. Our algorithm interfaces with CPLEX using python concert API of CPLEX.
Online embedding heuristic is implemented with python, using NetworkX library.

We use the fat tree topology  \cite{fares:2008} shown in (Figure~\ref{fig:fattree}) to represent the substrate data center.
For simplicity, we consider bandwidth of links and  memory capacity of switches as the attributes for the network.
Number of CPU cores and Memory are considered as the attributes of Servers and VMs.
% Number of nodes in the substrate graph are 128. 
VDC requests have random sized tree/fat tree topolgy - a most common case in big data processing clusters. The topologies were
generated using the topology generator of FNSS toolchain \cite{fnss}.
Simulation parameters are randomly distributed with ranges as as given in Table~\ref{tab:simvar}.
The arrivals of VDCN requests are determined by a Poisson distribution ranging from 1 to 10 requests per 100 time units.
% The bandwidth of links between core and aggregation switches are uniformly distributed between 100 to 1000 units.
% Whereas, that between servers, aggregation switches and edge switches 
% are uniformly distributed from  10 to 100 units. Node capacities are distributed between 50 and 100.
% % The virtual network topologies are generated using gt-itm tool.
% 
% The number of virtual nodes are uniformly distributed between 5 and 20.
% % We used various sizes of fat tree substrate networks for evaluation.

% The parameter $t^u_{dur}$ of virtual requests are uniformly distributed between 10 to 90 units of time.

% We consider the average waiting time of requests as one of the parameter for comparing the exact, heuristic and, our hybrid solution. 
% Figure~\ref{fig:wait} depicts the comparison of average waiting time  of the VDCN requests in our framework against the arrival rate.
% We compared the waiting time in our model against model similar to MIP model by Papagianni et al. \cite{papagianni:2013} \& and 
% heuristic similar to randomized heuristic by Rabbani et al. \cite{rabbani:2013}. 
% The figure shows that average waiting time of heuristic embedding is lower but, 
% HyViDE based strategy has waiting time comparable to that of heuristics.
% The waiting time of MIP based solution are high owing to the time spent by the requests before the batch embedding happens. 
% 
% \begin{figure}[t]
%  \centering
%   \includegraphics[width=.45\textwidth]{wait.eps}
%   \caption{Avg. waiting time for varying arrival rates}
%   \label{fig:wait}
% \end{figure}
% 

\begin{figure}[t]
 \centering
  \includegraphics[width=.45\textwidth]{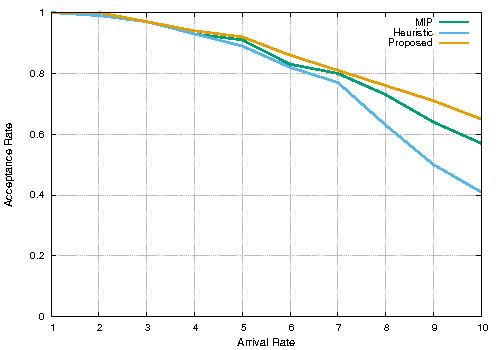}
  \caption{Acceptance rate for varying arrival rates}
  \label{fig:arr}
\end{figure}

The acceptance rate  with respect to varying rate of arrival is shown in Figure~\ref{fig:arr}. 
The figure clearly exposes the advantage of our strategies over MIP or heuristics solution used singly. 
Poor acceptance rate of heuristic can be accounted to the sub optimal solutions of heuristic methods. 
This leaves a lot of resources in data center unused. 
MIP solutions are applied in batches and hence the acceptance largely rely on the instantaneous availability of resources
whereas, the resources remain unused between successive embedding sessions.

\begin{figure}[t]
 \centering
  \includegraphics[width=.45\textwidth]{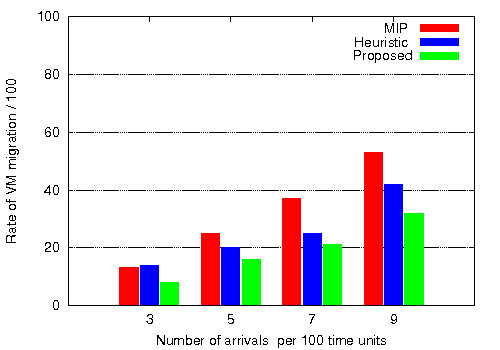}
  \caption{\% of VM migrations for different arrival rates}
  \label{fig:rate}
\end{figure}

Number of VM migrations  is a measure of quality of the VDCE solutions. Figure~\ref{fig:rate}
shows that our algorithm reduces the rate of VM migrations. Higher number of migrations in pure MIP solutions is intuitive from the nature of the algorithm. Every time a new batch is embedded,
the existing VMs get migrated with highest probability, to reach optimality. With pure MIP model it is hard to achieve optimal solution without VM migration in datacenter with fragmented resources. 
The remapping in heuristic algorithms are due to the periodic consolidation of resources, which is inevitable.

\algnewcommand{\algorithmicgoto}{\textbf{go to}}%
\algnewcommand{\Goto}[1]{\algorithmicgoto~\ref{#1}}%
\algrenewcommand\alglinenumber[1]{\tiny #1:}
\begin{algorithm}
\small
  \caption{Virtual Data Center Embedding}\label{alg:vne}
  \begin{algorithmic}[1]
    \Procedure{VDC\_Embed}{$G^s,\Re$}
%     \State $r\gets a\bmod b$
    \State $ts_c$ = residual server resource vector
    \State $ts_n$ = residual switch resource vector
    \State $ts_e$ = residual link resource vector
    \While{$|\Re|$ $\ne$ 0} \label{marker}
      \State Initialize the set $\Re '$ 
      \State Calculate the thresholds $(t^2_c, t^2_n,t^2_e)$ and $(t^1_c,t^1_n,t^1_e)$ 
      \If{$(ts_c,ts_n, ts_e) \ge (t^2_c, t^2_n,t^2_e)$)}
        \State call \textit{batch\_embedding($G^s,\Re '$)}
%         \State Add the non-embedded requests to $\Re$
      \ElsIf{($ (t^1_c, t^1_n,t^1_e) \le (ts_c, ts_n, ts_e) < (t^2_c, t^2_n,t^2_e)$)}
        \State Priority sort thet set $\Re$.
        \State Select $G^v_u \in \Re$ s.t. $t^u_{exp} \le t_{curr} + TD $ 
        \State call \textit{online\_embedding($G^s,G^v_u $)}
      \EndIf
    \EndWhile
   \EndProcedure
   \item[]   

   \Procedure{batch\_ embedding}{$G^s,\Re '$}
     \State Formulate the MIP (Section~\ref{sec:batch})
     \State Solve the MIP (with a solver)
     \For {VDC $u$ not embedded}
	\State Add $u$ back to $\Re$
     \EndFor
   \EndProcedure 
   \item[]
   \Procedure{online\_ embedding}{$G^s, G^v_u $}
     \If {residual\_ resource($C$) > requirement($G^v_u $)}
      \State Create temporary VDC placement $C$
      \State Confirm the embedding if possible with local search. 
     \Else 
      \State Select servers with least capacity violation.
      \State Find the $vi$ the amout of violations.
      \State Select smallest set of VMs with capacity $\ge vi$
      \State Swap with VM group where the difference \\in capacity is atleast $vi$
     \EndIf
   \EndProcedure
\end{algorithmic}
\end{algorithm}

\section{Conclusions} \label{sec:conclusion}
Our algorithm  is suitable for optimaly embedding virtual data centers on a physical data center.
The VDCE problem is important in placement of VMs with or without specific topology demands.
The proposed multi-mode algorithm maximizes the resource utilization over time by the online embedding, without waiting for embedding in batches. 
In case fragmentation is high, an MIP based batch re-embedding is applied to alleviate the same.
Sluggish ILP-based exact algorithm for batch embedding is the major limitation of our proposal. 
We are looking for faster mathematical programming techniques or near-optimal heuristics that can overcome the limitation. 
  
%\end{document}  % This is where a 'short' article might terminate

%ACKNOWLEDGMENTS are optional
% \section{Acknowledgments}
% The authors would like to thank Gerald Murray of ACM for
% his help in codifying this \textit{Author's Guide}
% and the \textbf{.cls} and \textbf{.tex} files that it describes.

% \bibliographystyle{apsrev4-1}
\bibliographystyle{plain}
\bibliography{mybib}  % sigproc.bib is the name of the Bibliography in this case

\begin{thebibliography}{10}

\bibitem{cplex}
Ilog cplex optimization studio v12.6.3, 2016.

\bibitem{fares:2008}
Mohammad Al-Fares, Alexander Loukissas, and Amin Vahdat.
\newblock A scalable, commodity data center network architecture.
\newblock {\em SIGCOMM Comput. Commun. Rev.}, 38(4):63--74, August 2008.

\bibitem{amokrane:2013}
A.~Amokrane, M.F. Zhani, R.~Langar, R.~Boutaba, and G.~Pujolle.
\newblock Greenhead: Virtual data center embedding across distributed
  infrastructures.
\newblock {\em IEEE Transactions on Cloud Computing}, 1(1):36--49, Jan 2013.

\bibitem{anderson:2002}
David~G Anderson.
\newblock Theoretical approaches to node assignment.
\newblock {\em Unpublished Manuscript}, 2002.

\bibitem{Belbekkouche2012}
Abdeltouab Belbekkouche, Md~Mahmud Hasan, and Ahmed Karmouch.
\newblock {Resource discovery and allocation in network virtualization}.
\newblock {\em IEEE Communications Surveys and Tutorials}, 14(4):1114--1128,
  2012.

\bibitem{chowdhury:2010}
Mosharaf Chowdhury, Fady Samuel, and Raouf Boutaba.
\newblock Polyvine: Policy-based virtual network embedding across multiple
  domains.
\newblock In {\em Proceedings of the Second ACM SIGCOMM Workshop on Virtualized
  Infrastructure Systems and Architectures}, VISA '10, pages 49--56, New York,
  NY, USA, 2010. ACM.

\bibitem{correa:2015}
E.S. Correa, L.A. Fletscher, and J.F. Botero.
\newblock Virtual data center embedding: A survey.
\newblock {\em Latin America Transactions, IEEE (Revista IEEE America Latina)},
  13(5):1661--1670, May 2015.

\bibitem{demchenko:2016}
Y.~Demchenko, F.~Turkmen, C.~de~Laat, C.~Blanchet, and C.~Loomis.
\newblock Cloud based big data infrastructure: Architectural components and
  automated provisioning.
\newblock In {\em 2016 International Conference on High Performance Computing
  Simulation (HPCS)}, pages 628--636, July 2016.

\bibitem{fischer:2013}
A.~Fischer, J.F. Botero, M.~Till~Beck, H.~de~Meer, and X.~Hesselbach.
\newblock Virtual network embedding: A survey.
\newblock {\em Communications Surveys Tutorials, IEEE}, 15(4):1888--1906,
  Fourth 2013.

\bibitem{greenberg:2009}
Albert Greenberg, James~R. Hamilton, Navendu Jain, Srikanth Kandula, Changhoon
  Kim, Parantap Lahiri, David~A. Maltz, Parveen Patel, and Sudipta Sengupta.
\newblock Vl2: A scalable and flexible data center network.
\newblock In {\em Proceedings of the ACM SIGCOMM 2009 Conference on Data
  Communication}, SIGCOMM '09, pages 51--62, New York, NY, USA, 2009. ACM.

\bibitem{haider:2009}
A.~Haider, R.~Potter, and A.~Nakao.
\newblock Challenges in resource allocation in network virtualization.
\newblock In {\em Proc. of 20th ITC Specialist Seminar on Network
  Virtualization}, Hoi An, Vietnam, 2009.

\bibitem{papagianni:2013}
C.~Papagianni, A.~Leivadeas, S.~Papavassiliou, V.~Maglaris, C.~Cervello-Pastor,
  and A.~Monje.
\newblock On the optimal allocation of virtual resources in cloud computing
  networks.
\newblock {\em IEEE Transactions on Computers}, 62(6):1060--1071, June 2013.

\bibitem{rabbani:2013}
M.G. Rabbani, R.~Pereira~Esteves, M.~Podlesny, G.~Simon,
  L.~Zambenedetti~Granville, and R.~Boutaba.
\newblock On tackling virtual data center embedding problem.
\newblock In {\em 2013 IFIP/IEEE International Symposium on Integrated Network
  Management (IM 2013)}, pages 177--184, Ghent, Belgium, May 2013.

\bibitem{rosa:2014}
R.~V. Rosa, C.~E. Rothenberg, and E.~Madeira.
\newblock Virtual data center networks embedding through software defined
  networking.
\newblock In {\em 2014 IEEE Network Operations and Management Symposium
  (NOMS)}, pages 1--5, Poland, May 2014.

\bibitem{fnss}
Lorenzo Saino, Cosmin Cocora, and George Pavlou.
\newblock A toolchain for simplifying network simulation setup.
\newblock In {\em Proceedings of the 6th International ICST Conference on
  Simulation Tools and Techniques}, SIMUTOOLS '13, Cannes, France, 2013. ICST.

\bibitem{sivaranjini:2015}
B.~Sivaranjani and P.~Vijayakumar.
\newblock A technical survey on various vdc request embedding techniques in
  virtual data center.
\newblock In {\em 2015 National Conference on Parallel Computing Technologies
  (PARCOMPTECH)}, pages 1--6, Bangalore,India, Feb 2015.

\bibitem{wang:2016}
T.~Wang, B.~Qin, and M.~Hamdi.
\newblock An efficient framework for online virtual network embedding in
  virtualized cloud data centers.
\newblock In {\em 2015 IEEE 4th International Conference on Cloud Networking
  (CloudNet)}, pages 159--164, Canada, Oct 2015.

\bibitem{Yu:2008}
Minlan Yu, Yung Yi, Jennifer Rexford, and Mung Chiang.
\newblock Rethinking virtual network embedding: Substrate support for path
  splitting and migration.
\newblock {\em SIGCOMM Comput. Commun. Rev.}, 38(2):17--29, March 2008.

\bibitem{qizhang:2014}
Qi~Zhang, M.F. Zhani, M.~Jabri, and R.~Boutaba.
\newblock Venice: Reliable virtual data center embedding in clouds.
\newblock In {\em 2014 Proceedings IEEE INFOCOM}, pages 289--297, Toronto,
  April 2014.

\bibitem{zhani:2013}
M.F. Zhani, Qi~Zhang, G.~Simon, and R.~Boutaba.
\newblock Vdc planner: Dynamic migration-aware virtual data center embedding
  for clouds.
\newblock In {\em 2013 IFIP/IEEE International Symposium on Integrated Network
  Management (IM 2013)}, pages 18--25, Ghent, Belgium, May 2013.

\end{thebibliography}
% ACM needs 'a single self-contained file'!
% \balancecolumns



%\bibliographystyle{IEEEtran}
% \end{thebibliography}

\end{document}